\documentclass[11pt]{article}
\usepackage[dvips,xdvi]{graphicx}

\begin{document}

\begin{titlepage}
\vspace{2cm}
\begin{center}
{\Large\bf Super-Kamiokande atmospheric neutrino results}\\[2cm]

{\large\bf Toshiyuki Toshito}\\
{\large for the Super-Kamiokande collaboration}\\

\vspace{1cm}

{\it Kamioka Observatory, Institute for Cosmic Ray Research,\\
 University of Tokyo, Japan.}

{\small \it present address: Department of physics, Nagoya university, Japan.} 
  
\end{center}

\vspace{2cm}

\begin{abstract}
We present atmospheric neutrino results from a 79 kiloton year (1289 days) exposure of the Super-Kamiokande detector.
Our data are well explained by $\nu_{\mu} \rightarrow \nu_{\tau}$
2-flavor oscillations.
%We update the 3 flavor analysis of contained events. 
We have been attempting to discriminate between the
possible oscillating partners of the muon neutrino as being either
the tau neutrino or the sterile neutrino.
%These tests use expected differences due to neutral currents
%and matter effects to discriminate the possibilities.
%We find no evidence favoring sterile neutrinos, and reject the hypothesis at
%99\% confidence level.
Pure $\nu_\mu\rightarrow\nu_s$ oscillation is disfavored at 99\% C.L..
Moreover, we performed the appearance search for charged current tau neutrino
interactions in the upward-going samples.
Our data is consistent with $\nu_{\tau}$ appearance
at roughly the two-sigma level. 

\end{abstract}

\vspace{2cm}
\begin{center}
  Talk presented on XXXVIth Rencontres de Moriond\\
  Electroweak Interactions and Unified Theories\\
  10-17 March 2001.
\end{center}

\end{titlepage}

\pagestyle{empty}

\section{Introduction}
Atmospheric neutrino are produced as decay products in hadronic
showers resulting from collisions of cosmic-rays with nuclei in the upper
atmosphere.
Production of electron and muon neutrino is dominated by the processes
$
\pi^\pm\rightarrow\mu^\pm+\nu_\mu(\bar{\nu_\mu})
$
followed by
$
\mu^\pm\rightarrow e^\pm+\bar{\nu_\mu}(\nu_\mu)+\nu_e(\bar{\nu_e}).
$
That gives an expected ratio of the flux of $\nu_{\mu}$ to the flux of
$\nu_{e}$ of about 2.
Vertically downward-going neutrinos travel about 15km
while vertically upward-going neutrinos travel about 13,000km before
interacting in the detector.
Thanks to good geometrical symmetry of the earth,
we can expect up-down symmetry of neutrino flux.
Details in prediction of atmospheric neutrino flux is discussed on  
Ref. 1.
%\cite{Honda:1995}.
Neutrino oscillation occurs if a finite mass difference and
mixing angle exists.
The oscillation probability between two neutrino flavors is given by 
$P = \sin^22\theta\cdot
\sin^2(1.27\frac{L(\mathrm{km})}{E_\nu(\mathrm{GeV})}\Delta m^2(\mathrm{eV}^2))$,
where $\theta$ is the mixing angle, $L$ is the
flight length of the neutrino, $E_\nu$ is the neutrino energy,
$\Delta m^2$ is the mass squared difference.
The range of energy of observable atmospheric neutrino is from
a few hundred MeV to the order of 100GeV.
The broad energy spectrum and flight distances makes measurement of
atmospheric neutrino sensitive to neutrino oscillation with $\Delta m^{2}$
down to the order of $10^{-4}\mathrm{eV}^{2}$.
Several recent underground experiments report atmospheric neutrino
results in terms of neutrino oscillation\cite{Fukuda:1998mi,Ambrosio:1998wu,Kafka:1999kk}.  
This paper reports on recent results of the Super-Kamiokande.

\begin{table*}[btp]
\caption{Event summary for 1289 days contained sample}\label{tab:evsum}
%\begin{tabular}{cc}
%\begin{minipage}[]{0.40\textwidth }
\begin{tabular}{c|cc}
\multicolumn{3}{l}{sub-GeV ($E_{vis}<1.33\mathrm{GeV}$)} \\
 & Data & MC(Honda flux) \\ \hline
1ring $e$-like & 2864 & 2668 \\
1ring $\mu$-like & 2788 & 4073 \\ \hline
multi ring & 2159 & 2585 \\ \hline
Total & 7811 & 9326 \\ \hline
\multicolumn{3}{l}{} \\
\multicolumn{3}{l}{} \\
\multicolumn{3}{l}{} \\
\multicolumn{3}{l}{}
\end{tabular}
%\end{minipage}
%&
%\begin{minipage}[]{0.40\textwidth }
\begin{tabular}{c|cc}
\multicolumn{3}{l}{multi-GeV FC($E_{vis}>1.33\mathrm{GeV}$)} \\
 & Data & MC(Honda flux) \\ \hline
1ring $e$-like & 626 & 613 \\
1ring $\mu$-like & 558 & 838 \\ \hline
multi ring & 1318 & 1648 \\ \hline
Total & 2502 & 3099 \\ \hline
\multicolumn{3}{l}{} \\
\multicolumn{3}{l}{Partially Contained} \\ \hline
Total & 754 & 1065 \\ \hline
\multicolumn{3}{l}{}
\end{tabular}
%\end{minipage}
%\end{tabular}
\end{table*}

\section{Event sample}
\begin{figure}[tbp]
\begin{center}
    \includegraphics[width=0.8\textwidth]{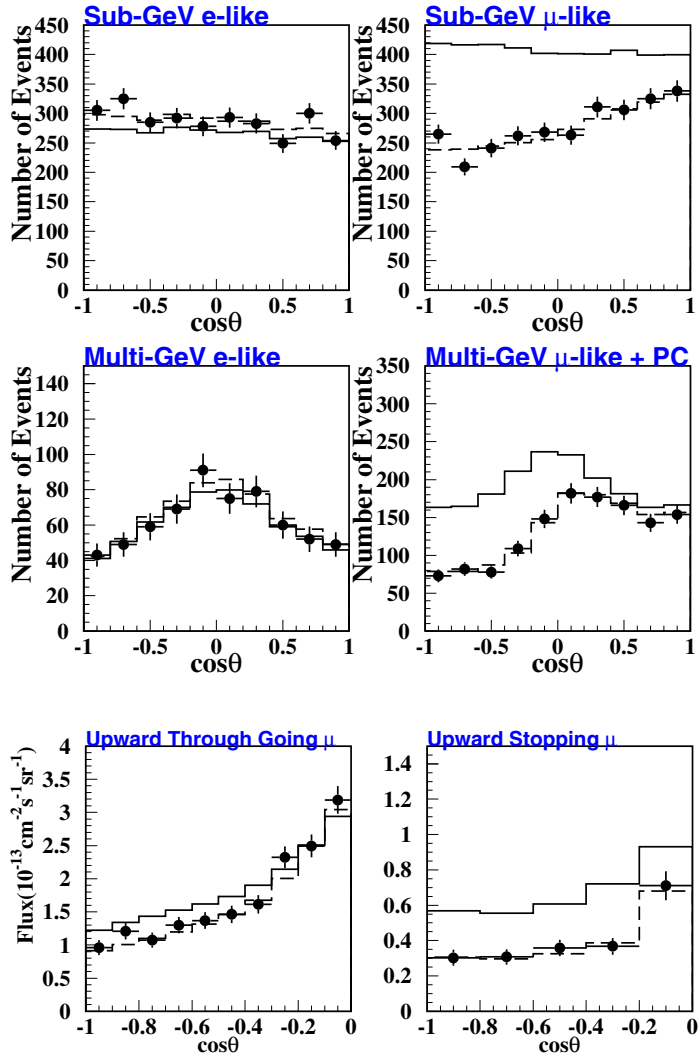}

%\epsfile{file=comb_zenith.eps,width=0.5\textwidth}

\end{center}
\caption{Zenith angle distribution of Super-Kamiokande 1289 days
FC, PC and UPMU samples. Dots, solid line and dashed line correspond
to data, MC with no oscillation and MC with best fit oscillation parameters,
respectively.}
\label{fig:zen_all}
\end{figure}

Super-Kamiokande is a 50 kiloton water
Cherenkov detector constructed under Mt.
Ikenoyama located at the central part of Japan, giving it a 
rock over-burden of 2,700 m water-equivalent. 
The fiducial mass of the detector for atmospheric neutrino
analysis is 22.5 kiloton.
Neutrino events interacting with the water are observed 
as fully-contained (FC)
or partially-contained (PC) events according to the amount of 
anti-counter activity.
FC events with only one reconstructed ring are
subdivided into $e$-like and $\mu$-like based
on likelihood analysis of the reconstructed Cherenkov ring.
Our 1289 live-days data are summarized
in Table~\ref{tab:evsum} and their zenith angle distributions
are shown in Figure~\ref{fig:zen_all}. 
The flavor ratio is evaluated by taking the double ratio using MC
expectation without oscillation as:
$
R\equiv\frac{(\mu\mathrm{-like}/e\mathrm{-like})_\mathrm{DATA}}{(\mu\mathrm{-like}/e\mathrm{-like})_\mathrm{MC}}.
$
The observed values are $0.638^{+0.017}_{-0.017}\pm 0.050$
for sub-GeV and $0.675^{+0.034}_{-0.032}\pm 0.080$ for multi-GeV
samples. 
These small double ratio is consistent with our previous result\cite{Fukuda:1998mi}
and indecates a deficit of muon neutrino explained
by neutrino oscillation. 
Neutrino events interacting with rock surrounding the detector can be
observed as upward through going muons or stopping muons.
Details in the analysis can be found on 
Ref. 2.
%\cite{Fukuda:1998mi}.

\section{$\nu_\mu\rightarrow\nu_\tau$ oscillation analysis}
\begin{figure*}
%\begin{tabular}{cc}
%\begin{minipage}{0.5\textwidth }
\begin{center}
%\epsfile{file=contour.1d.comb.eps,height=7cm}
%\epsfile{file=cont1d_ichep.eps,height=7cm}
%    \includegraphics[width=0.8\textwidth]{cont1d_ichep.eps}
    \includegraphics[width=0.8\textwidth]{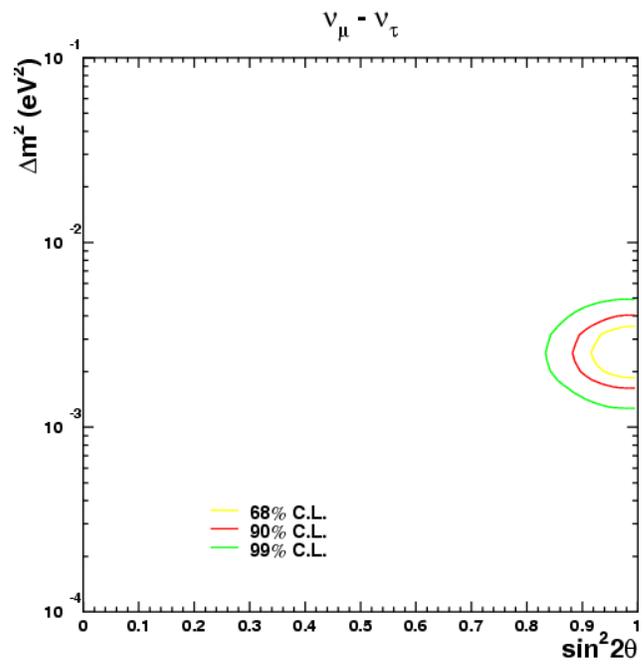}
\end{center}
\caption{
68,90 and 99\% confidence level allowed regions for $\nu_\mu\rightarrow \nu_\tau$
oscillation obtained by Super-Kamiokande 1289 days result.
}
\label{fig:contour}
\end{figure*}

Allowed oscillation parameter region using contained and upward going muon
samples is shown
in Fig.~\ref{fig:contour}.
The minimum $\chi^2$ including unphysical region is found to be 142.1 with 152 degrees of
freedom (d.o.f.) at $\Delta m^2 = 2.5\times 10^{-3} \mathrm{eV}^2$,
$\sin^22\theta = 1.00$. 
The deficit of upward going $\mu$-like data is well
explained by assuming $\nu_\mu\rightarrow\nu_\tau$ oscillation (Figure~\ref{fig:zen_all}).
$\chi^2$ for no oscillation was found to be 344.1 for 154 d.o.f. 

\section{Study on $\nu_\mu\rightarrow\nu_\tau$ and $\nu_\mu\rightarrow\nu_{sterile}$}
\begin{figure*}[btp]
\begin{center}
%\epsfile{file=tau-s_zenith.eps,width=1.0\textwidth}
    \includegraphics[width=1.0\textwidth]{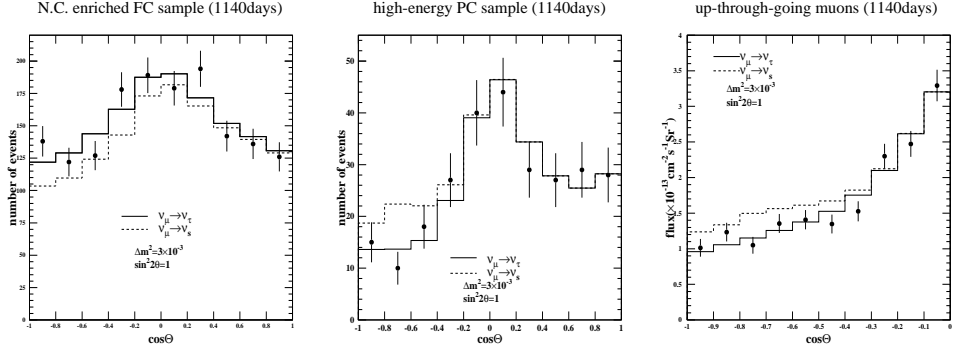}
\end{center}
\caption{Zenith angle distributions of left: NC enriched sample,
center: high-energy PC sample, right: up-through-going muon sample.
}
\label{fig:zen_tau_s}
\end{figure*}

\begin{figure}[ht]
\begin{center}
%\epsfile{file=tau-s_cont.eps,width=0.7\textwidth}
    \includegraphics[width=1.2\textwidth]{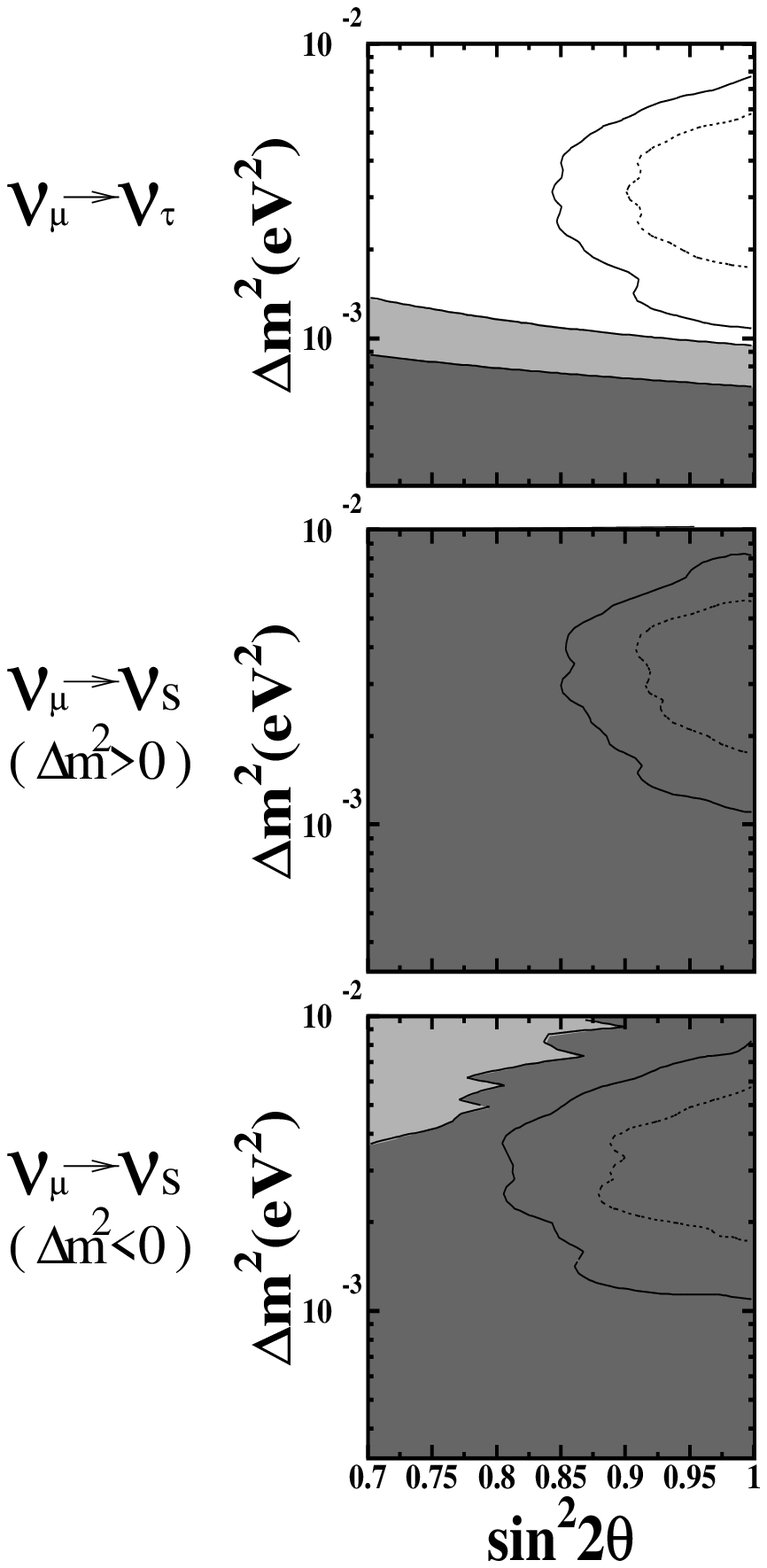}
\end{center}
\caption{Excluded regions for three oscillation modes.
The light(dark) gray region
is excluded at 90(99)\% C.L.
by NC enriched sample and high-energy sample analysis.
Thin dotted(solid) line indicates the
90(99)\% C.L. allowed regions from 1ring-FC sample analysis.
}
\label{fig:contour_tau_s}
\end{figure}

Some models predict
that $\nu_\mu$ oscillates into ``sterile'' neutrino ($\nu_s$) that
does not interact even via neutral current (NC). 
If the observed deficit of $\nu_\mu$ is due to $\nu_\mu\rightarrow
\nu_s$ oscillation, then the number of events produced via 
NC interaction for up-going neutrino
should also be reduced. 
Moreover, in the case of $\nu_\mu\rightarrow\nu_s$ oscillation, matter effect will
suppress oscillation in the high energy ($E_\nu > 15$GeV) region\cite{matter}.
We used the following data sample to observe these effects:
(a) NC enriched sample; (b) the high-energy 
($E_{vis}$ $>$ 5GeV) PC sample; and 
(c) up-through-going muons.  
Zenith angle distributions for each sample is shown in 
Fig.~\ref{fig:zen_tau_s}.
The hypothesis test is performed using the
up($\cos\Theta<-0.4$)/down($\cos\Theta>0.4$) ratio in samples (a) and (b) and 
the vertical($\cos\Theta<-0.4$)/horizontal($\cos\Theta>-0.4$) ratio in sample (c).
Fig.~\ref{fig:contour_tau_s} shows excluded regions obtained by combined
((a),(b)and(c)) analysis
along with the allowed regions
from 1ring-FC sample analysis 
assuming $\nu_\mu\rightarrow\nu_\tau$ and $\nu_\mu\rightarrow\nu_s$.
The results show that
pure $\nu_\mu\rightarrow\nu_s$ oscillation is disfavored at 99\%. 

\section{Search for charged current $\nu_{\tau}$}

We performed appearance search for charged current (CC) $\nu_{\tau}$.
In this analysis we assumed that
2 flavor $\nu_\mu\rightarrow\nu_\tau$ oscillations happen
at $\Delta m^2 = 3\times 10^{-3}eV^2$ and $\sin^22\theta = 1$.
%In this assumption about 20 CC $\nu_\tau$ events/year
%are expected at Super-Kamiokande.
The difference between CC $\nu_{\tau}$ and others
appear in the energy spectrum, number of charged pions in the
final state, fraction of missing energy with respect to 
neutrino energy and so on. 
We have done three different analysis to enrich
CC $\nu_\tau$.
The first one is likelihood method using parameters as visible energy
,number of ring ,number of decay electron and so on.
The second one is neural network method using similar parameters
which are used in the first analysis.
The third analysis is likelihood method using parameters as energy flow
and event shape.
$\nu_\tau$ events appear as upward going events.
Our analyses are optimized by MC
and looking at only downward going events in data in which
no signals are expected in order to perform blind analyses.
CC $\nu_\tau$ events are observed as the excess of
upward-going events in the zenith angle distributions
if we consider only the contribution of $\nu_{e}$,$\nu_{\mu}$ and NC.
Results of there analyses are summarized in Table~\ref{tab:tausum}.
%According to MC,
%74 CC $\nu_\tau$ events are expected so far.
MC predicts 74 CC $\nu_\tau$ events so far.
Each result of three analyses and MC expectation
agrees well within the statistical error.
%Our data is consistent with $\nu_\tau$ appearance.
Zero $\nu_\tau$ appearance is disfavored at roughly the two-sigma level.
The three analyses are highly correlated with each
other, and thus cannot be combined in an independent way to increase the
significance.

\begin{table*}[btp]
\caption{Summary of three analyses}\label{tab:tausum}
\begin{tabular}{l|ccc}
analysis & 1 & 2 & 3 \\ \hline
number of CC $\nu_{\tau}$ 
& $43 \pm 17 ^{+8}_{-11}$ 
& $44 \pm 20 ^{+8}_{-12}$
& $25.5 ^{+14}_{-13}$ \\ \hline
efficiency
& 0.42 & 0.45 & 0.32 \\ \hline
number of CC $\nu_{\tau}$ (efficiency corrected) 
& $103 \pm 41 ^{+18}_{-26}$ 
& $98 \pm 44 ^{+18}_{-27}$
& $79 ^{+44}_{-40}$ \\ \hline
\end{tabular}
\end{table*}

\section{Summary}
Super-Kamiokande results from 1289 days
of contained
events and up-going muons events give
90\% C.L. allowed parameter regions of
$\sin^22\theta >0.88$ and 
$1.6\times 10^{-3} <\Delta m^2 < 4\times 10^{-3}\mathrm{eV}^2$.
%The results of 3-flavor oscillation analysis
%is consistent with the CHOOZ and
%2-flavor $\nu_\mu\rightarrow\nu_\tau$ oscillation.
%Finally,
Pure $\nu_\mu\rightarrow\nu_s$ oscillation is disfavored at 99\% C.L..
According to the results of search for charged current $\nu_\tau$ events,
our data is consistent with $\nu_\tau$ appearance
at roughly the two-sigma level.

%%%%%%%%%%%%%%%%%%%%%%%%%

\end{document}